\documentclass[a4paper,11pt]{amsart}

\usepackage{amsthm,amscd,amsmath,amssymb}
\usepackage[all]{xy}

\newcommand{\squishlist}{
   \begin{list}{$\bullet$}
    { \setlength{\itemsep}{0pt}      \setlength{\parsep}{3pt}
      \setlength{\topsep}{3pt}       \setlength{\partopsep}{0pt}
      \setlength{\leftmargin}{1.5em} \setlength{\labelwidth}{1em}
      \setlength{\labelsep}{0.5em} } }

\newcommand{\squishlisttwo}{
   \begin{list}{$\bullet$}
    { \setlength{\itemsep}{0pt}    \setlength{\parsep}{0pt}
      \setlength{\topsep}{0pt}     \setlength{\partopsep}{0pt}
      \setlength{\leftmargin}{2em} \setlength{\labelwidth}{1.5em}
      \setlength{\labelsep}{0.5em} } }

\newcommand{\squishend}{
    \end{list}  }

\newtheorem{theorem}{Theorem}[section]
\newtheorem{lemma}[theorem]{Lemma}

\newtheorem{proposition}[theorem]{Proposition}

\theoremstyle{definition}
\newtheorem{definition}[theorem]{Definition}
\theoremstyle{remark}
\newtheorem{remark}[theorem]{Remark}
\newtheorem{example}[theorem]{Example}
\newcommand{\coloneq}{\mathrel{\mathop:}=}

\DeclareMathOperator{\dom}{dom}
\DeclareMathOperator{\codom}{codom}
\DeclareMathOperator{\Ar}{Ar}
\DeclareMathOperator{\Rep}{Rep}
\DeclareMathOperator{\WRep}{WRep}
\DeclareMathOperator{\Mod}{Mod}
\DeclareMathOperator{\LMod}{LMod}

\DeclareMathOperator{\Mon}{Mon}

\newcommand{\rar}{\longrightarrow}
\newcommand{\LC}{{\mathsf{LC}}}
\newcommand{\var}{{\mathsf{var}}}
\newcommand{\app}{{\mathsf{app}}}
\newcommand{\abs}{{\mathsf{abs}}}

\newcommand{\Cat}[1]{\mathsf{#1}}
\newcommand{\CC}{\Cat{C}}
\newcommand{\DD}{\Cat{D}}
\newcommand{\Set}{{\mathsf{Set}}}
\newcommand{\eval}{{\mathsf{eval}}}
\newcommand{\subst}{{\mathsf{subst}}}
\newcommand{\swap}{{\mathsf{swap}}}

\title{Higher-order theories}
\author{Andr\'e Hirschowitz \and Marco Maggesi}

\begin{document}
\maketitle

\begin{abstract}
  We extend our approach to abstract syntax (with binding
  constructions) through modules and linearity. First we give a new
  general definition of arity, yielding the companion notion of
  signature. Then we obtain a modularity result as requested by
  \cite{GU03}: in our setting, merging two extensions of syntax
  corresponds to building an amalgamated sum.  Finally we define a
  natural notion of equation concerning a signature and prove the
  existence of an initial semantics for a so-called
  \emph{representable} signature equipped with a set of equations.
\end{abstract}

\section{Introduction}

\subsection{Modules for modular higher-order syntax and semantics}

Many programming or logical languages allow constructions which bind
variables and this higher-order feature causes much trouble in the
formulation, the understanding and the formalization of the theory of
these languages. For instance, there is no universally accepted
discipline for such formalizations: that is precisely why the POPLmark
Challenge \cite{PoplMark} offers benchmarks for testing old and new
approaches. Although this problem may ultimately concern typed
languages and their operational semantics, it already concerns untyped
languages equipped with an equational semantics. Indeed, even at the
informal level, there is not yet a universally adopted notion of
higher-order theory.

The goal of this work is to extend, in particular towards equational
semantics, our approach to higher-order abstract syntax \cite{HM},
based on modules and linearity.  First of all, we give a new general
definition of arity, yielding the companion notion of signature. The
notion is coined in such a way to induce a companion notion of
representation of an arity (or of a signature) in a monad: such a
representation is a morphism among modules over the given monad, so
that an arity simply assigns two modules to each monad. Then we
explain how our approach enjoys modularity in the sense introduced by
\cite{GU03}: in our view, a syntax is or yields a monad equipped with
some module morphisms; such enriched monads form a category in which
merging two extensions of syntax corresponds to building an
amalgamated sum. Finally, we define a general notion of equation for a
given signature (or syntax) $\Sigma$ and we prove the corresponding
theorem for initial semantics.  As for arities, our notion of equation
is coined in such a way to induce the notion of whether the equation
is satisfied by an arbitrary representation of $\Sigma$ in a monad
$R$: this will be when two morphisms of modules over $R$
coincide. Accordingly an equation assigns to each representation of
$\Sigma$ a pair of morphisms of modules (with common domain and
codomain).

\subsection{Examples}

Such a proposal should definitely offer a convincing picture
concerning the paradigmatic example of the lambda-calculus.  In our
previous work \cite{HM} we treated the examples of the untyped
lambda-calculus modulo $\alpha$-equivalence and
$\alpha\beta\eta$-equivalence.  In section \ref{sec:examples} we
revisit these two examples. We then
give various other examples obtained by adding features to the lambda-calculus:
differentiation, parallelism, explicit substitution.

\subsection{Future work}

In the present work we give a notion of untyped higher-order theory,
and we observe the existence of an initial representation for
\emph{algebraic} signatures.  This has to be extended to more general
signatures, including the signature of the lambda-calculus with
explicit substitution, as pioneered in \cite{GU03}, in order to reach
a satisfactory state for the untyped equational setting.  The point of
view proposed here will also be accommodated to deal with languages
with a fixed set of types (for an example see \cite{Zsido}), or to
model operational semantics through monads on the category of
preordered sets, or both.  It should also be extended to the case of
dependent types.

\subsection{Organization of the paper}

Section \ref{sec:modules} gives a succinct account about modules over
a monad.  Our new definitions of (higher-order) arity and signature are
given in section \ref{sec:higher-order}.  We propose a solution to the
problem of modularity in section \ref{sec:modularity}.  Section
\ref{sec:half-equations} develops the theory of equations for our
notion of syntax and contains our theorem about initial semantics.
Section \ref{sec:examples} treats some explicit examples.  The last
section discusses related works.

\section{Modules over Monads}
\label{sec:modules}

We recall only the definition and some basic facts about (right)
modules over a monad.  See \cite{HM} for a more extensive introduction
on this topic.

Let $\CC$ be a category.  A monad over $\CC$ is a monoid in the
category $\CC\to\CC$ of endofunctors of $\CC$, i.e., a triple
$R=\langle R,\mu,\eta \rangle$ given by a functor $R\colon\CC \to
\CC$, and two natural transformations $\mu\colon R^2 \to R$ and
$\eta\colon I \to R$ such that the following diagrams commute:
\begin{equation*}
  \xymatrix{
    R^3 \ar[d]_{\mu R} \ar[r]^{R\mu} & R^2 \ar[d]^{\mu} \\
    R^2 \ar[r]^{\mu} & R}\qquad
  \xymatrix{
    I\cdot R \ar[rd]_{1_R} \ar[r]^{\eta R} & R^2 \ar[d]^{\mu} &
    R\cdot I \ar[ld]^{1_R} \ar[l]_{R \eta}  \\
    {} & R {}}
\end{equation*}

Let $R$ be a monad over $\CC$.

\begin{definition}[Right modules]
  A right $R$-module is given by a functor $M\colon \CC \to \DD$
  equipped with a natural transformation $\rho\colon M \cdot R \to M$,
  called \emph{action}, which is compatible with the monad
  composition and identity:
  \begin{equation*}
    \xymatrix{
      M\cdot R^2 \ar[d]_{\rho R} \ar[r]^{M\mu} & M\cdot R \ar[d]^{\rho} \\
      M\cdot R \ar[r]^{\rho} & M}\qquad
    \xymatrix{
      M\cdot R \ar[d]^{\rho} &
      M\cdot I \ar[ld]^{1_M} \ar[l]_{M \eta}  \\
      M}
  \end{equation*}
  We will refer to the category $\DD$ as the \emph{range} of $M$.

  We say that a natural transformation of right $R$-modules
  $\tau\colon M \to N$ is \emph{linear} if it is compatible with
  action:
  \begin{equation*}
    \xymatrix{
      M\cdot R \ar[d]_{\rho_M} \ar[r]^{\tau R} & N\cdot R \ar[d]^{\rho_N} \\
      M \ar[r]^{\tau} & N}
  \end{equation*}
  We take linear natural transformations as morphisms among right
  modules having the same range $D$.  It can be easily verified that
  we obtain in this way a category that we denote $\Mod^\DD(R)$.
\end{definition}

There is an obvious corresponding definition of left $R$-modules that
we do not need to consider in this paper.  From now on, we will write
$R$-modules instead of right $R$-modules for brevity.

\begin{example}
  \label{ex:modules}
  Let us show some trivial examples of modules:
  \begin{enumerate}
  \item Every monad $R$ is a module over itself, which we call
    the \emph{tautological} module.
  \item For any functor $F\colon \DD \to \Cat{E}$ and any $R$-module
    $M\colon \CC \to \DD$, the composition $F\cdot M$ is a $R$-module
    (in the evident way).
  \item For every object $W\in \DD$ we denote by $\underbar W \colon
    \CC \to \DD$ the constant functor $\underbar W \coloneq X \mapsto
    W$.  Then $\underbar W$ is trivially a $R$-module since
    $\underbar W = \underbar W \cdot R$.
  \end{enumerate}
\end{example}

Limits and colimits in the category of right modules can be
constructed point-wise.  For instance:
\begin{lemma}[Limits and colimits of modules]
  If $\DD$ is complete (resp.\ cocomplete), then $\Mod^\DD(R)$ is
  complete (resp.\ cocomplete).
\end{lemma}

In particular, we will often make use of the fact that, if the range
category $\DD$ is cartesian, then the category $\Mod^\DD(R)$ is also
cartesian.

For our purposes, one important example of module is given by the
following general construction.  Let $\CC$ be a category with finite
colimits and a final object $*$.
\begin{definition}[Derivation]
  For any $R$-module $M$ with range $\DD$, the \emph{derivative} of
  $M$ is the functor $M \coloneq X \mapsto M(X+*)$.  Derivation can be
  iterated, we denote by $M^{(k)}$ the $k$-th derivative of $M$.
\end{definition}

\begin{proposition}
  Derivation yields an endofunctor of $\Mod^\DD(R)$.  Moreover, if
  $\DD$ is a cartesian category, derivation is a cartesian endofunctor
  of $\Mod^\DD(R)$.
\end{proposition}

In the case $\CC=\DD=\Set$, the functor $M'$ is given by $M'\coloneq X
\mapsto M(X+*)$, where $X+*$ denotes the set obtained by adding a new
point to $X$.  Moreover, we have a natural \emph{evaluation
  morphism}
\begin{equation*}
  \eval \colon M'\times R \rar M.
\end{equation*}
which is $R$-linear.  This allows us to interpret the derivative $M'$
as the ``module $M$ with one formal parameter added''.  Higher-order
derivatives have analogous morphisms (that we still denote with
$\eval$)
\begin{equation*}
  \eval \colon M^{(b)}\times R^b \rar M.
\end{equation*}
where $\eval(t,m_1,\dots,m_b) \in M(X)$ is obtained by substituting
$m_1, \dots, m_b \in R(X)$ in the successive stars of $t\in M^{(b)}(X)
= M(X + * + \dots + *)$.

We already introduced the category $\Mod^\DD(R)$ of modules with fixed
base $R$ and range $D$.  It it often useful to consider a larger
category which collects modules with different bases.  To this end, we
need first to introduce the notion of pull-back.

\begin{definition}[Pull-back]
  Let $f\colon R\to S$ be a morphism of monads and $M$ a $S$-module.
  The action
  \begin{equation*}
    M\cdot R \stackrel{Mf}\rar M \cdot S \stackrel\rho\rar M
  \end{equation*}
  defines a $R$-module which is called \emph{pull-back} of $M$ along
  $f$ and noted $f^*\!M$.  It can be easily verified that a $S$-linear
  natural transformation $g\colon M \to N$ is also a $R$-linear
  natural transformation $f^*g \colon f^*M \to f^* N$ and that
  $f^*\colon \Mod^\DD(S) \to \Mod^\DD(R)$ is a functor.
\end{definition}

It can be easily verified that pull-back is well-behaved with respect
to many important constructions.  In particular:
\begin{proposition}
  \label{prop:pull-back-commute}
  Pull-back commutes with products and with derivation.
\end{proposition}

\begin{definition}\label{dfn:arity-morph}
  Given a list of non negative integers $(a)=(a_1,\dots,a_n)$ we
  denote by $M^{(a)}=M^{(a_1,\dots,a_n)}$ the module
  $M^{(a_1)}\times\cdots\times M^{(a_n)}$.  Observe that, when
  $(a)=()$ is the empty list, we have $M^{()}=*$ the final module.
\end{definition}

\begin{definition}[The large module category]
  We define the large module category $\LMod$ as follows:
  \squishlist %% \begin{itemize}
  \item its objects are pairs $(R, M)$ of a monad $R$ and a $R$-module
    $M$.
  \item a morphism from $(R, M)$ to $(S, N)$ is a pair $(f, m)$ where
    $f\colon R \rar S$ is a morphism of monads, and $m\colon M \rar
    f^*N$ is a morphism of $R$-modules.  The category $\LMod$ comes
    equipped with a forgetful functor to the category of monads, given
    by the projection $(R,M) \mapsto R$.
  \squishend %% \end{itemize}
\end{definition}

\section{Half-arities and signatures}
\label{sec:higher-order}

In this section, we improve our approach to higher-order syntax
\cite{HM}, and give a new notion of arity. The destiny of an arity is
to have representations in monads. A representation of an arity $a$ in
a monad $R$ will be a morphism between two modules $\dom(a, R)$ and
$\codom(a, R)$.  For instance, in the case of the arity $a$ of
$\app_2$, we have $\dom(a, R)\coloneq R^2$ and $\codom(a, R)\coloneq
R$. Hence an arity consists of two halves, each of which assigns to
each monad $R$ a module over $R$ in a functorial way.

\begin{definition}[Arity]
  A \emph{half-arity} is a right-inverse functor to the projection
  from the category $\LMod$ to the category $\Mon$ of monads.  An
  \emph{arity} is a pair of two half-arities. The arity $(a,b)$ is
  denoted $a \to b$; $a$ and $b$ are called respectively the domain
  and the codomain of $a \to b$.
\end{definition}

\begin{example}
\begin{enumerate}
\item The assignment $R \mapsto R$ is a half-arity which we denote by
  $\Theta$.
\item The assignment $R \mapsto *$, where $*$ denotes the final module
  over $R$ is a half-arity which we denote by $*$.
\item Given a half-arity $a$, for each non-negative integer $n$, the
  assignment $R \mapsto a(R)^{(n)}$ is a half-arity which we denote by
  $a^{(n)}$.  As usual, we also set $a' \coloneq a^{(1)}$ and
  $a''\coloneq a^{(2)}$.
\item Given two half-arities $a$ and $b$, the assignment $R \mapsto
  a(R)\times b(R)$ is a half-arity which we denote by $a \times b$ .
\item For each non-negative integer $n$, the assignment $R \mapsto R^{n}$ is a
  half-arity which we denote by $\Theta^{n}$.
\item For each sequence of non-negative integers $s$, the assignment $R \mapsto
  R^{(s)}$ is a half-arity which we denote by $\Theta^{(s)}$.
\item The assignment $R \mapsto R \cdot R$ is a half-arity which we
  denote by $\Theta \cdot \Theta$.  Of course we have plenty of such
  composite half-arities.
\end{enumerate}
\end{example}

We denote by $\Ar$ the set of arities.

\begin{definition}[Algebraic and raw arities]
  Half-arities of the form $\Theta^{(s)}$ and arities of the form
  $\Theta^{(s)} \to \Theta^{(t)}$ are said algebraic.  An arity of the
  form $H \to \Theta$ is said raw.
\end{definition}

These algebraic arities are slightly more general than those in
\cite{FPT}, which are precisely algebraic raw arities. In particular
we have an algebraic arity ($\Theta \to \Theta'$) for the $\app_1$
construction given in section \ref{sec:lambda-calculus-alpha}.

\begin{definition}[Signatures]
  We define a signature $\Sigma =(O, \alpha)$ to be a family of
  arities $\alpha \colon O \to \Ar$.  A signature is said to be raw
  (resp. algebraic) if it consists of raw (resp. algebraic)
  arities.
\end{definition}

\begin{definition}[Representation of an arity, of a signature]
  Given a monad $R$ over $\Set$, we define a representation of the
  arity $a$ in $R$ to be a module morphism from $\dom(a, R)$ to
  $\codom (a, R)$; a representation of a signature $S$ in $R$ consists
  of a representation in $R$ for each arity in $S$.
\end{definition}

\begin{example}
  The usual $\app\colon \LC^2 \to \LC$ (see section
  \ref{sec:lambda-calc-alphabetaeta}) is a representation of $\Theta^2
  \to \Theta$ into $\LC$.  A representation of $* \to \Theta''$ in
  $\LC$ is given by the $\app_0$ construction ($\app_0 (X) \coloneq
  \app (\var(*_{X+*}), \var(*_X))$).
\end{example}

\begin{definition}[The category of representations]
 Given a signature $\Sigma = (O, \alpha)$, we build the category
 $\Mon^{\Sigma}$ of representations of $\Sigma$ as follows.  Its
 objects are monads equipped with representations of $\Sigma$.  A
 morphism from $(M, r)$ to $(N, s)$ is a morphism from $M$ to $N$
 compatible with the representations in the sense that, for each $o$
 in $O$, the following diagram
 commutes:
 \begin{equation}
   \begin{CD}
     \dom (\alpha(o), M) @>>> \codom (\alpha(o), M) \\
     @VVV          @VVV \\
     f^* \dom (\alpha(o), N)  @>>> f^* \codom (\alpha(o), N)
   \end{CD}
 \end{equation}
 where the horizontal arrows come from the representations and the
 vertical arrows come from the functoriality of half-arities.
\end{definition}

\begin{proposition}
 These morphisms, together with the obvious composition, turn
 $\Mon^\Sigma$ into a category which comes equipped with a forgetful
 functor to the category of monads.
\end{proposition}

\begin{definition}
 \label{def:representable}
 A signature $\Sigma$ is said representable if the category
 $\Mon^\Sigma$ has an initial object.
\end{definition}

\begin{theorem}
 \label{thm:syntax}
  Algebraic signatures are representable.
\end{theorem}

For more details we refer to our paper \cite{HM} (theorems 1 and 2).

\begin{remark}
  There is a slightly more general notion of arity and signature.
  Roughly speaking, a $\Sigma$-arity will be a pair of
  $\Sigma$-modules (see below section 5).  Such a $\Sigma$-arity may
  be added to $\Sigma$, yielding a bigger signature.  This picture
  allows to consider partially defined constructions, like the
  predecessor.  We leave this extension for future work.
\end{remark}

\section{Modularity}
\label{sec:modularity}

It has been stressed \cite{GU03} that the standard approach (via
algebras) to higher-order syntax lacks modularity.  In the present
section we show in which sense our approach via modules enjoys
modularity.

Suppose that we have a signature $\Sigma = (O, a)$ and two
subsignatures $\Sigma'$ and $\Sigma''$ covering $\Sigma$ in the
obvious sense, and let $\Sigma_0$ be the intersection of $\Sigma'$ and
$\Sigma''$.  Suppose that these four signatures are representable (for
instance because $\Sigma$ is algebraic).  Modularity would mean that
the corresponding diagram of monads
\begin{equation*}
  \xymatrix{
    \hat \Sigma_0\ar[r]\ar[d] & \hat\Sigma'\ar[d]\\
    \hat \Sigma''\ar[r] & \hat\Sigma}
\end{equation*}
is cocartesian.  The observation of \cite{GU03} is that the diagram
of raw monads is, in general, not cocartesian.  Since we do not want
to change the monads, in order to claim for modularity, we will have
to consider a category of enriched monads.  Here by enriched monad, we
mean a monad equipped with some additional structure.

Our solution to this problem goes through the following category $\WRep$:
\squishlist %% \begin{itemize}
\item An object of $\WRep$ is a triple $(R, \Sigma, r)$ where $R$ is a
  monad, $\Sigma$ a signature, and $r$ is a representation of $\Sigma$ in
  $R$.
\item A morphism in $\WRep$ from $(R, (O, a), r)$ to $(R', (O', a'),
  r')$ consists of a map $i\coloneq O \to O'$ compatible with $a$ and
  $a'$ and a morphism $m$ from $(R, r)$ to $(R', i^*(r'))$, where, for
  $i$ injective, $i^*(r')$ should be understood as the restriction of
  the representation $r'$ to the subsignature $(O, a)$.
\item It is easily checked that the obvious composition turns $\WRep$
  into a category.
\squishend %% \end{itemize}
Now for each signature $\Sigma$, we have an obvious functor from
$\Mon^\Sigma$ to $\WRep$, through which we may see $\hat \Sigma$ as an
object in $\WRep$.  Furthermore, an injection $i\colon \Sigma_1 \to
\Sigma_2$ obviously yields a morphism $i_*\coloneq \hat \Sigma_1 \to
\hat \Sigma_2$ in $\WRep$.  Hence our `cocartesian' square of
signatures as described above yields a square in $\WRep$. The proof of
the following statement is straightforward.

\begin{proposition}
  Modularity holds in $\WRep$, in the sense that given a `cocartesian'
  square of signatures as described above, the associated square in
  $\WRep$ is cocartesian again.
\end{proposition}

There is a stronger statement in a category $\Rep$ with the same
objects and more morphisms.  Roughly speaking, in $\Rep$, a morphism
from $(R, \Sigma, r)$ to $(R', \Sigma', r')$ is a compatible pair of a
monad morphism and a ``vertical'' functor from $Mon^{\Sigma'}$ to
$Mon^{\Sigma}$.  We plan to describe this more carefully in some
future work.

\section{The Category of Half-Equations}
\label{sec:half-equations}

In this section, we are given a signature $\Sigma$, and we build the
category where our equations will live.

\begin{definition}[The category of half-equations]
  We define a $\Sigma$-module $U$ to be a functor from the category of
  representations of $\Sigma$ to the category $\LMod$ commuting with the
  forgetful functors to the category $\Mon$ of monads.
  \begin{equation*}
    \xymatrix{
      \Mon^\Sigma \ar[rr]^U\ar[dr] & & \LMod \ar[dl]\\
      & \Mon}
  \end{equation*}
  We define a morphism of $\Sigma$-modules to be a natural
  transformation which becomes the identity when composed with the
  forgetful functor. We call these morphisms ``half-equations''.

  These definitions yield a category which we call the category of
  $\Sigma$-modules or half-equations.
\end{definition}

\begin{example}
  To each half-arity $a$ is associated, by composition with the
  projection from $Mon^\Sigma$ to $Mon$, a $\Sigma$-module still
  denoted $a$.  In particular we have a $\Sigma$-module $\Theta$.
  Accordingly, to each construction $c \in \Sigma$ with arity $a \to
  b$ is associated in the obvious way a morphism of $\Sigma$-modules,
  still denoted $c$, from $a$ to $b$.
\end{example}

\begin{example}
  Since derivation is a "vertical" endofunctor in $\LMod$, it acts on
  $\Sigma$-modules.  In particular we also have a family of
  $\Sigma$-modules $\Theta^{(n)}$.
\end{example}

\begin{proposition}
  The category of half-equations is cartesian.
\end{proposition}

\begin{example}
  To each application $f\colon [1, \dots, p] \rar [1, \dots, q]$ is
  associated a half-equation: $f_*\colon T^{(p)} \rar T^{(q)}$ (which
  we call renaming along $f$).
\end{example}

\begin{definition}[Equations]
  We define an equation for $\Sigma$ to be a pair of half-equations
  with common source and target.  We also write $e_1=e_2$ for the
  equation $(e_1, e_2)$.
\end{definition}

\begin{example}
  In case $\Sigma$ consists of the two constructions $\abs$ and
  $\app_1$ (cfr.~section \ref{sec:lambda-calculus-alpha}) with
  respective arities $\Theta' \to \Theta$ and $\Theta \to \Theta'$,
  the $\beta$ equation is $\app_1 \cdot\abs = \text{Id}_{\Theta'}$,
  while the $\eta$ equation is $\abs \cdot\app_1 = \text{Id}_\Theta$.
\end{example}

\begin{definition}[Satisfying equations]
  We say that a representation $r$ of $\Sigma$ in a monad $M$
  satisfies the equation $e_1 = e_2$ if $e_1(r)=e_2(r)$.  If $E$ is a
  set of equations for $\Sigma$, we say that a representation $r$ of
  $\Sigma$ in a monad $M$ satisfies $E$ (or is a representation of
  $(\Sigma, E)$) if it satisfies each equation in $E$.  We define the
  category of representations of $(\Sigma, E)$ to be the full
  subcategory in the category of representations of $\Sigma$ whose
  objects are representations of $(\Sigma, E)$.
\end{definition}

\begin{theorem}[Initial representation of $(\Sigma, E)$]
  Given a set $E$ of equations for a representable signature $\Sigma$,
  the category of representations of $(\Sigma, E)$ has an initial
  object.
\end{theorem}

\begin{proof}[Sketch]
  We build the monad $S$ for our initial representation as a quotient
  of $\hat \Sigma$ (see theorem \ref{thm:syntax}).  For each set $X$,
  we define an equivalence relation $r_X$ on $\hat \Sigma (X)$ as
  follows: $r_X(a, b)$ means that for any representation $\rho$ of
  $(\Sigma, E)$ in a monad $M$, $i_X(a)$ equals $i_X(b)$, where
  $i\colon \hat \Sigma \rar M$ is the initial functor associated to
  $\rho$.  We check easily that this is an equivalence relation, that
  the corresponding collection of quotients inherits the structure of
  monad, and that this quotient monad has the required universal
  property.
\end{proof}

\section{Examples}
\label{sec:examples}

We now illustrate the previous notions through some well-known
examples.

\subsection{Monoids}
\label{sec:monoids}

We first consider an example of first-order syntax with equations.
Given a set $X$, let us denote by $M(X)$ the free monoid built over
$X$.  This is a classical example of monad over the category of
(small) sets.  The monoid structure gives us, for each set $X$, two
maps $m_X\colon M(X) \times M(X) \to M(X)$ and $e_X\colon * \to M(X)$
given by the product and the identity respectively.  It can be easily
verified that $m\colon M^2 \rar M$ and $e\colon * \to M$ are
$M$-linear natural transformations.  In other words $(M,\rho) = (M,\{
m, e \})$ constitutes a representation of the signature $\Sigma = \{
m\colon\Theta^2\to \Theta, e\colon * \to \Theta\}$.

In the category $\Mon^\Sigma$ of representations of $\Sigma$ we
consider the \emph{associativity equation}, i.e., the pair of
half-equations given by
\begin{equation*}
  \xymatrix@C=25pt@R=2pt{
    \Theta^3 \ar[rr]^{\Theta\times m} &&
    \Theta^2\ar[r]^{m} & \Theta \\
    \Theta^3 \ar[rr]_{m\times \Theta} &&
    \Theta^2\ar[r]_{m} & \Theta}
\end{equation*}
Analogously, we define the left and right \emph{identity equations}
given by the morphisms
\begin{equation*}
  \xymatrix@C=25pt@R=2pt{
    \Theta\ar[r]^{e\times \Theta} & \Theta^2 \ar[r]^{m} & \Theta \\
    \Theta \ar[rr]_{=} && \Theta}
  \quad\text{and}\quad
  \xymatrix@C=25pt@R=2pt{
    \Theta\ar[r]^{\Theta\times e} & \Theta^2 \ar[r]^{m} & \Theta \\
    \Theta\ar[rr]_{=} && \Theta}
\end{equation*}
and we denote by $E$ the system constituted by these three equations.

The category $\Mon^{(\Sigma,E)}$ is the category of monads with a
structure of monoid and $(M,\rho)$ is its initial object.

\subsection{Lambda-calculus modulo $\alpha$-equivalence}
\label{sec:lambda-calculus-alpha}

We denote by $\Lambda(X)$ the set of lambda-terms up to
$\alpha$-equivalence with free variables ``indexed'' by the set $X$.
It is well-known \cite{BPdebruijn,Alt-Reus,HM} that $\Lambda$ has a
natural structure of monad where the monad composition is given by
variable substitution.

It can be easily verified (\cite{HM}) that application and abstraction
are $\Lambda$-linear natural transformations
\begin{equation*}
  \app\coloneq \Lambda^2 \rar \Lambda,\qquad
  \abs\coloneq \Lambda' \rar \Lambda.
\end{equation*}
that is, $\Lambda$ is a monad endowed with a representation $\rho$ of
the signature $\Sigma = \{\app\colon\Theta^2\to \Theta,
\abs\colon\Theta'\to \Theta\}$.

Again, the monad $\Lambda$ is initial in the category $\Mon^\Sigma$ of
monads endowed representations of the signature $\Sigma$.

For our purposes, it is important to introduce another interesting,
non-raw, representation on $\Lambda$ which is defined as follows.
Consider the $\Lambda$-module morphism
\begin{equation*}
  \app_1\colon \Lambda \rar \Lambda'
\end{equation*}
given by
$\app_1(x) = \app(x,*)$.
Then $\app_1$ is a representation of arity $\Theta \to \Theta'$ and
the usual $\app$ constructor can be recovered from $\app_1$ by
\emph{flattening}, i.e.,
\begin{equation*}
  \app(x,y) = \eval(\app_1(x),y).
\end{equation*}
Then we can consider on $\Lambda$ the representation $\rho' =
\{\app_1,\abs\}$ of signature $\Sigma' = \{\app_1\colon\Theta\to \Theta',
\abs\colon\Theta'\to \Theta\}$.  The categories $\Mon^{\Sigma'}$ and
$\Mon^\Sigma$ are equivalent through the \emph{flattening} and
$(\Lambda,\rho')$ is the initial object of the category
$\Mon^{\Sigma'}$.

\subsection{Lambda-calculus modulo $\alpha\beta\eta$-equivalence}
\label{sec:lambda-calc-alphabetaeta}

Now we can introduce the lambda-calculus modulo
$\alpha\beta\eta$-equivalence as a quotient of the previous calculus.
We denote by $\LC(X)$ the set of lambda-terms up to
$\alpha\beta\eta$-equivalence with free variables ``indexed'' by the
set $X$.  As in the previous example, the monad $\LC$ is endowed with
a representation $\rho$ and $\rho'$ of the two signatures $\Sigma = \{
\app\colon\Theta^2 \to \Theta, \abs\colon\Theta' \to \Theta \}$ and
$\Sigma' = \{ \app_1\colon\Theta \to \Theta', \abs_1\Theta' \to \Theta
\}$ respectively.  Now we want to introduce the equations for the
$\beta$ and $\eta$ equivalence relations.  This can be done both with
$\rho$ and $\rho'$, but it looks more natural when it is formulated
with respect to $\rho'$.

We define the $\beta$ and $\eta$ equivalence relation as the pair of
half-equations on $\Mon^{\Sigma'}$ given by
\begin{equation*}
  \xymatrix@R=2pt{
    \Theta' \ar[r]^{\abs} &\Theta \ar[r]^{\app_1} &\Theta' \\
    \Theta' \ar[rr]_{=} && \Theta'}
  \quad\text{and}\quad
  \xymatrix@R=2pt{
    \Theta \ar[r]^{\app_1} &\Theta' \ar[r]^{\abs} &\Theta \\
    \Theta \ar[rr]_{=} && \Theta}
\end{equation*}
respectively.  We denote by $E$ the system constituted by the $\beta$
and $\eta$ equations.  It can be shown that $(\LC,\rho')$ is initial
in the category $\Mon^{(\Sigma',E)}$ (this has been proved formally in
the Coq proof assistant and discussed in \cite{HM}).

\subsection{HOcore}
\label{sec:line-algebr-lambda}

$\mathsf{HOcore}$ is a higher-order calculus for concurrency
introduced by Lanese, P{\'e}rez, Sangiorgi and Schmitt
\cite{Lanese2008On-the-Expressivenes}.  The syntax of HOcore is as
follows
\begin{equation*}
  P\quad ::=\quad x\quad |\quad a(x).P\quad |
            \quad \bar{a}.\langle P\rangle\quad |\quad P\ ||\ P\quad |
            \quad \mathbf{0}
\end{equation*}
where $a$ ranges over the sort of \emph{names} (or \emph{channels})
and $x$ denote a process variable.  In the input prefix process
$a(x).P$ the variable $x$ is bound in the body $P$.  In our framework
the $\mathsf{HOcore}$ syntax is represented by the signature
\begin{align*}
  \Sigma_{\mathsf{HOcore}} =
    \{\quad &\forall a.\ \mathsf{recv}_a\colon \Theta' \to \Theta,\quad
       \forall a.\ \mathsf{send}_a\colon \Theta \to \Theta,\\
       &\mathsf{parallel}\colon \Theta^2 \to \Theta,\quad
       \mathsf{zero}\colon * \to \Theta \quad\}
\end{align*}
where $\mathsf{recv}$ and $\mathsf{send}$ are families of constructors
parametrized over the names $a$.  Let us denote by $\mathsf{HOcore}$
the monad $\hat\Sigma_{\mathsf{HOcore}}$.

To spell out a concrete example, consider a pair of concurrent
processes where the first process sends a message $P$ through the
channel $a$ to the second process:
\begin{equation*}
  \bar{a}\langle P\rangle\ ||\ a(x).Q
\end{equation*}
In our formalism such process is represented by the syntax tree
\begin{equation*}
  \mathsf{parallel} (\mathsf{send}_a P, \mathsf{recv}_a Q)
\end{equation*}
where $P\in \mathsf{HOcore}(X)$ and $Q\in \mathsf{HOcore}(X+*)$ for
some set of name variables $X$.  Now consider the reduction rule
\begin{equation*}
  \bar{a}\langle P\rangle\ ||\ a(x).Q\quad \equiv\quad
  \mathbf{0}\ ||\ Q[x\coloneq P]
\end{equation*}
which models the exchange of a message $P$ through a channel $a$
between two concurrent processes.  This can be stated as the equality
of the two linear morphisms of kind
\begin{math}
  \mathsf{HOcore}\times\mathsf{HOcore}' \rar \mathsf{HOcore}
\end{math}
given by
\begin{align*}
  (P,Q) &\mapsto \mathsf{parallel} (\mathsf{send}_a P, \mathsf{recv}_a Q) \\
  (P,Q) &\mapsto \mathsf{parallel} (\mathsf{zero}, \eval(Q, P))
\end{align*}
Once abstracted over the representation $\rho$, the previous pair of
morphisms gives a pair of half-equations on the category
$\Mon^{\Sigma_\mathsf{HOcore}}$.

\subsection{Lambda calculus with explicit substitution}
\label{sec:lambda-calculus-explicit}

We now consider an example of non algebraic signature.  On the monad
$\Lambda(X)$ of lambda calculus (modulo $\alpha$-equivalence) given in
\ref{sec:lambda-calculus-alpha} the substitution operator
\begin{equation*}
  \subst \colon \Lambda\cdot\Lambda \rar \Lambda
\end{equation*}
given by the monad composition (or \emph{join}) is a representation of
the arity $\Theta\cdot\Theta \to \Theta$ ($\Theta\cdot\Theta$ has a
natural structure of module thanks to \ref{ex:modules} (2)).  We then
have a representation $\rho$ of the signature
\begin{equation*}
  \Sigma \coloneq \{\quad
  \app\colon\Theta^2\to \Theta,\quad
  \abs\colon\Theta'\to \Theta,\quad
  \subst\colon\Theta\cdot\Theta \to \Theta\quad
  \}.
\end{equation*}
In the monad $\Lambda$, substitution interacts with the $\app$ and
$\abs$ constructors in the following way.  Let us denote by $\subst'
\colon \Theta\cdot\Theta'=(\Theta\cdot\Theta)' \to \Theta'$ the linear
transformation induced by $\subst$.  Given any monad $P$ we have a
natural transformation $\swap_X \colon P(X)+* \rar P(X+*)$.  Then
$(\Lambda, \rho)$ satisfies the system $E$ constituted by the two
equations
\begin{equation*}
  %\label{eq:explicit-app}
  \xymatrix@C=30pt@R=2pt{
    \Theta\cdot \Theta\times \Theta\cdot \Theta\ar[rr]^{\app\cdot\Theta} &&
    \Theta\cdot \Theta \ar[r]^\subst & \Theta\\
    \Theta\cdot \Theta\times \Theta\cdot \Theta\ar[rr]_{\subst\times\subst} &&
    \Theta\times \Theta \ar[r]_\app & \Theta}
\end{equation*}
and
\begin{equation*}
  %\label{eq:explicit-abs}
  \xymatrix@R=2pt{
    \Theta'\cdot \Theta\ar[rr]^{\abs\cdot\Theta} &&
    \Theta\cdot \Theta \ar[r]^\subst & \Theta\\
    \Theta'\cdot \Theta\ar[r]_{\Theta\cdot\swap} & \Theta\cdot \Theta' \ar[r]_{\subst'} &
    \Theta'\ar[r]_\abs & \Theta}
\end{equation*}
We call $\Mon^\Sigma$ the category of lambda-calculi with \emph{explicit
substitution}.   We claim that $(\Lambda, \rho)$ is its initial
object in $\Mon^{(\Sigma,E)}$.

\subsection{Lambda-calculus with explicit differentiation}
\label{sec:lambda-calculus-diff}

Inspired by \cite{ER03} we sketch the definition of what we call
\emph{Lambda-calculus with explicit differentiation}
($\mathsf{LCED}$).  The syntax of $\mathsf{LCED}$ is given by the
signature
\begin{align*}
  \Sigma_{\mathsf{LCED}} \coloneq \{\quad
    &\app \colon \Theta^2\to \Theta, \quad
    \abs \colon \Theta' \to \Theta, \quad
    \mathbf{0} \colon * \to \Theta, \quad
    + \colon \Theta^2\to\Theta, \\
    &\mathsf{partial} \colon \Theta'\times\Theta\to \Theta', \quad
    \forall k\in\mathbb{N}^*.\,\mathsf{diff}_k \colon \Theta^2\to \Theta
  \quad\}.
\end{align*}
We use the notations $\partial t \cdot u$ for $\mathsf{partial}(u, t)$
(the ``linear'' substitution --- in the sense of \cite{ER03} --- of $u$
for the first variable in $t$) and $D_k t \cdot u$ for
$\mathsf{diff}_k(t, u)$ (the differential of $t$ with respect to the
$k$-th variable applied to $u$).

The link between differentiation and linear substitution is
essentially given by the rule
\begin{equation*}
  D_1(\abs\, t)\cdot u = \abs\, (\partial t \cdot u)
\end{equation*}
The explicit linear substitution has its set of associated rules.
E.g., for the substitution of an application we have
\begin{equation*}
  \partial(\app(s,t))\cdot u = \app(\partial s\cdot u, t) +
  \app( D_1 s\cdot (\partial t \cdot u), t ).
\end{equation*}

We omit the complete set of rules of the calculus which will be
treated in a future work.

\section{Related Works}
\label{sec:related}

The idea that the notion of monad is suited for modelling substitution
concerning syntax (and semantics) has been retained by many recent
contributions concerned with syntax (see e.g.\
\cite{BPfold,GU03,MU03}) although some other settings have been
considered.  Notably in \cite{FPT} the authors work within a setting
roughly based on operads (although they do not write this word down).
Our main specificity here is the systematic use of the observation
that the natural transformations we deal with are linear with respect
to natural structures of module (a form of linearity had already been
observed, in the operadic setting, see \cite{FT01}, section 4).

The signatures we consider here are slightly more general than the
signatures in \cite{FPT}, which allows us to give a signature to our
$\app_1$. On the other hand, our signatures `reduce' (by
\emph{flattening}) to those in \cite{FPT}.  In some future work, we
plan to recover signatures as in \cite{MU03} in terms of modules and
to extend our initial semantics to this setting.

\bibliographystyle{amsalpha}
\bibliography{lc}

\end{document}